\documentclass[a4paper]{jpconf}
\usepackage{graphicx,amsmath,amsfonts,amssymb, dcolumn,amsthm}
\begin{document}
\title{Induced gravity from gauge theories}

\author{R.~F.~Sobreiro and A.~A.~Tomaz}

\address{UFF $-$ Universidade Federal Fluminense, Instituto de F\'{\i}sica, Campus da Praia Vermelha, Avenida General Milton Tavares de Souza s/n, 24210-346, Niter\'oi, RJ, Brasil.}

\ead{sobreiro@if.uff.br and tomaz@if.uff.br}

\author{V.~J.~Vasquez Otoya}

\address{IFSEMG $-$ Instituto Federal de Educa\c{c}\~ao, Ci\^encia e Tecnologia, Rua Bernardo Mascarenhas 1283, 36080-001, Juiz de Fora, MG, Brasil.}

\ead{victor.vasquez@ifsudestemg.edu.br}

\begin{abstract}
We discuss the possibility of a class of gauge theories, in four Euclidean dimensions, to describe gravity at quantum level. The requirement is that, at low energies, these theories can be identified with gravity as a geometrodynamical theory.  Specifically, we deal with de Sitter-type groups and show that a Riemann-Cartan first order gravity emerges. An analogy with quantum chromodynamics is also formulated. Under this analogy it is possible to associate a soft BRST breaking to a continuous deformation between both sectors of the theory, namely, ultraviolet and infrared. Moreover, instead of hadrons and glueballs, the physical observables are identified with the geometric properties of spacetime. Furthermore, Newton and cosmological constants can be determined from the dynamical content of the theory.
\end{abstract}

\section{Introduction}\label{intro}

Since the advent of general relativity (GR) \cite{Einstein:1916vd}, gravity has been showing itself to be very different from all other fundamental interactions. The three fundamental interactions contemplated by the standard model are quantum mechanically stable gauge theories. In fact, quantum field theory (QFT) \cite{Itzykson:1980rh} describes fluctuating fields which are parameterized by a set of rigid spacetime coordinates $X^\mu$ while spacetime itself is a Minkowski space (To perform actual computations, spacetime must be Wick-rotated to the Euclidean space). On the other hand, GR is a geometrodynamical theory \cite{Hojman:1976vp} that describes spacetime as a dynamical entity, \emph{i.e.}, spacetime coordinates gain a dynamical character. If one declare that the metric tensor $g_{\mu\nu}(X)$ is the fundamental field of gravity, a standard quantization is automatically forbidden because its quantum nature would be transferred to spacetime. For instance, if $\hat{g}_{\mu\nu}(X)$ is the quantum version of the metric tensor then, $\hat{X}_\mu=\hat{g}_{\mu\nu}(X)X^\nu$, which implies on $\hat{X}^\mu=\hat{g}^{\mu\nu}(X)\hat{X}_\nu$. Thus, spacetime coordinates would acquire a quantum character. The inconsistency rises because a quantum field cannot be parameterized by ``quantum coordinates''. Axiomatic QFT states that spacetime coordinates must be a rigid set of well-defined parameters \cite{Streater:1989vi}. Essentially, geometrodynamics appears to be an exclusively classical description of gravity. 

Another property that ruins a quantum description of gravity is the success of Einstein equations in describing macroscopic phenomena. At macroscopic scale, the \emph{rhs} of Einstein equations is the energy-momentum tensor of the matter distribution, which is an effective object. Thus, the \emph{lhs} of Einstein equations should also be an effective object and not a fundamental tensor that could be easily quantized by standard techniques.

To avoid the incompatibility between the principles of QFT and GR, the main method adopted is to expand the metric tensor around a background Minkowski metric $g_{\mu\nu}=\eta_{\mu\nu}+kh_{\mu\nu}$, where $k$ is related to Newton's constant and carry ultraviolet dimension $-1$. The perturbation $h_{\mu\nu}$ is interpreted as the graviton field. However, it is widely known that the linearized Einstein-Hilbert (EH) action, known as the Pauli-Fierz action \cite{Fierz:1939ix}, does not define a stable action at quantum level \cite{'tHooft:1974bx,Deser:1974cz,Deser:1974cy}. It is necessary then to introduce higher derivative terms which, however, destroy the unitarity of the theory \cite{Stelle:1976gc}.

To get gravity closer to the rest of the fundamental interactions, it is convenient to write it as a gauge theory \cite{Utiyama:1956sy,Kibble:1961ba,Sciama:1964wt}. This description of gravity is known as the first order formalism (The formalism based on the metric tensor is called second order formalism), see also \cite{Mardones:1990qc,Zanelli:2005sa}. In this approach, gravity is described by two fundamental 1-form fields, the vierbein\footnote{The indices conventions are: $\{\mathfrak{a},\mathfrak{b},\mathfrak{c},\ldots\}\in\{0,1,2,3\}$ are frame indices; and $\{\mathrm{greek}\}\in\{0,1,2,3\}$ are world indices.} $e^\mathfrak{a}$ and the spin-connection ${\omega^\mathfrak{a}}_\mathfrak{b}$. The geometric properties of spacetime are obtained from specific gauge invariant composite fields \cite{Sobreiro:2010ji,Sobreiro:2010qf}. In particular, the metric tensor is obtained from
\begin{equation}
g_{\mu\nu}=\eta_\mathfrak{ab}e^\mathfrak{a}_\mu e^\mathfrak{b}_\nu\;,\label{geom0}
\end{equation}
while the affine connection is determined by
\begin{equation}
\Gamma^\alpha_{\mu\nu}=g^{\alpha\beta}\eta_\mathfrak{ab}\left(e^\mathfrak{b}_\beta\partial_\mu e^\mathfrak{a}_\nu+e^\mathfrak{b}_\beta\omega^\mathfrak{a}_{\mu\mathfrak{c}}e^\mathfrak{c}_\nu\right)\;.\label{geom1}
\end{equation}
The vierbein is a map between coordinates $X^\mu$ at a point $X$, in spacetime manifold $\mathbb{M}^4$, and coordinates $X^a$, in the tangent space $T_X(\mathbb{M})$, at the very same point $X$, \emph{i.e.}, $dX^\mathfrak{a}=e^\mathfrak{a}_\mu dX^\mu$. The spin-connection is related to parallel transport between near tangent spaces $T_X(\mathbb{M})$ and $T_{X+dX}(\mathbb{M})$ and is recognized as the gauge field of the theory. Typically, the gauge group is the Lorentz group $SO(1,3)$ and describes the local isometries of spacetime. Nevertheless, since the fundamental fields are still related to spacetime properties, any attempt of standard quantization of these fields would lead to the same inconsistency that occur in the second order formalism. Another attempt to solve the QFT-GR incompatibility is to declare that the fundamental fields are not immediately related to spacetime. Instead, the geometric properties of spacetime emerge as effective phenomena \cite{Zanelli:2005sa,Sobreiro:2010ji,Sobreiro:2010qf}. For instance,
\begin{eqnarray}
g_{\mu\nu}&=&\eta_\mathfrak{ab}\left<e^\mathfrak{a}_\mu e^\mathfrak{b}_\nu\right>\;,\nonumber\\
\Gamma^\alpha_{\mu\nu}&=&g^{\alpha\beta}\eta_\mathfrak{ab} \left<e^\mathfrak{b}_\beta\partial_\mu e^\mathfrak{a}_\nu+e^\mathfrak{b}_\beta\omega^a_{\mu \mathfrak{c}}e^\mathfrak{c}_\nu\right>\;.\label{geom2}
\end{eqnarray}
The idea is to work with geometric variables with no specific relation to the metric tensor nor the affine connection. Although promising, it is unclear to work with a ill defined coordinate system.

Sticking to the gauge theoretical approach, many other theories have been proposed by generalizing the gauge groups and their respective actions. In particular \cite{Stelle:1979va,Stelle:1979aj,Pagels:1983pq,Tresguerres:2008jf,Mielke:2010zz}, it is worth mentioning the de Sitter groups $SO(m,n)$, with $(m+n)=5$, in four-dimensional spacetime, supplemented with a spontaneous symmetry breaking mechanism \cite{Higgs:1964pj,Higgs:1964ia}. In the work \cite{MacDowell:1977jt}, gravity and supergravity are considered under the framework of a dynamical breaking instead of a spontaneous breaking. In any of these cases, the Yang-Mills action \cite{Yang:1954ek} is not considered. When the Yang-Mills action is taken into account, the starting theory is already metric dependent, see for instance \cite{Tseytlin:1981nu,Mahato:2004zi}. Only in \cite{Sobreiro:2007pn}, the Yang-Mills was considered independently of the metric. The Lorentz group was taken as the gauge group of a theory in flat spacetime. A relation with a Palatini-type gravity emerges from a color symmetry breaking generated by a condensation mechanism.

More recently \cite{Sobreiro:2011hb}, the Yang-Mills action for de Sitter groups was studied independently of the geometric properties of spacetime. Essentially, asymptotic freedom \cite{Gross:1973id,Politzer:1973fx}, mass parameters and an In\"on\"u-Wigner contraction \cite{Inonu:1953sp} to Lorentz-type groups account for the emergence of a gravity theory. The absence of mass parameters in the starting action is important because it prevents the identification of the gauge field with the vierbein. Thus, it is important that the masses are dynamically generated. One possibility is the Gribov parameter \cite{Gribov:1977wm,Zwanziger:1992qr,Sobreiro:2005ec} which is required for quantum consistency at low energy scales. Moreover, the presence of the Gribov parameter induces a soft BRST breaking \cite{Baulieu:2008fy,Baulieu:2009xr,Dudal:2012sb}, which continuously deforms the massless theory in flat spacetime into a gravity theory in the first order formalism. Extra dynamical masses can also contribute to the model \cite{Dudal:2005na,Dudal:2011gd}. Remarkably, Newton and cosmological constants can be determined from these masses and the Yang-Mills coupling parameter. Thus, since pure Yang-Mills theory is renormalizable \cite{'tHooft:1972fi,Piguet:1995er}, this model can be regarded as a possible theory for quantum gravity.

In the present work, we exploit some physical properties of the theory developed in \cite{Sobreiro:2011hb}. In particular, we discuss the analogy between this theory and quantum chromodynamics (QCD), the soft BRST breaking, the dynamical masses contribution and the possibility of the theory in providing reliable physical results.

Section 2 is devoted to the foundations of Yang-Mills theories, BRST symmetry and confinement. Section 3 is dedicated to a QCD-Gravity analogy. In Section 4, the details of de Sitter gauge theories in four-dimensional Euclidean spacetime and their relation with gravity are discussed. In Section 5, some consistency checks of the model are provided. Our final considerations can be found in Section 6.

\section{Yang-Mills theories, BRST symmetry and confinement}

\subsection{Preliminary concepts}\label{gauge}

The so called Yang-Mills theories \cite{Yang:1954ek} are gauge theories based on a gauge symmetry described by a semi-simple Lie group $G$. This class of theories consists of a generalization of electrodynamics, which is an Abelian gauge theory for the group $U(1)$. The fundamental field is the gauge connection 1-form $Y=Y^{\mathsf{a}}J_{\mathsf{a}}$, where $J_{\mathsf{a}}$ are the generators of the group and the group indices vary as $\{\mathsf{a},\mathsf{b},\mathsf{c},\ldots\}\in\{1,2,\ldots,\dim{G}\}$. The curvature 2-form is defined as $F=\nabla^2=\mathrm{d}Y+\kappa YY$, where $\nabla=\mathrm{d}+\kappa Y$ is the covariant derivative, $\mathrm{d}$ is the exterior derivative in spacetime and $\kappa$ is the coupling parameter.

The Yang-Mills action is constructed as
\begin{equation}
S_{\mathrm{YM}}=\frac{1}{2}\int\;F^{\mathsf{a}}\ast F_{\mathsf{a}}\;,\label{ym0}
\end{equation}
where $\ast$ is the Hodge operator. The action \eqref{ym0} is invariant under gauge transformations,
\begin{equation}
Y\longmapsto g^{-1}\left(\frac{1}{\kappa}\mathrm{d}+Y\right)g\;,\label{gt0}
\end{equation}
where $g\in G$. To define a path integral for the Yang-Mills action, a gauge fixing must be imposed \cite{Itzykson:1980rh,Faddeev:1967fc}. For a consistent introduction of a gauge fixing, it is convenient to adopt the BRST quantization method \cite{Piguet:1995er,Tyutin:1975qk,Becchi:1975nq}. For simplicity, we adopt the Landau gauge, $\mathrm{d}\ast Y=0$. In this framework, the action \eqref{ym0} is replaced by
\begin{equation}
S_0=S_{\mathrm{YM}}+\int\left(ib^\mathsf{a}\mathrm{d}\ast Y_\mathsf{a}+\overline{c}_\mathsf{a}\mathrm{d}\ast\nabla^\mathsf{ab}c_\mathsf{b}\right)\;,\label{ym1}
\end{equation}
where the fields $c^\mathsf{a}$ and $\overline{c}^\mathsf{a}$ are the ghost and anti-ghost fields, respectively, and $b^\mathsf{a}$ is a Lagrange multiplier which enforces the Landau gauge. The components of the covariant derivative are $\nabla^\mathsf{ab}=\delta^\mathsf{ab}\mathrm{d}-\kappa f^\mathsf{abc}Y_\mathsf{c}$, where $f^\mathsf{abc}$ are the structure constants of the gauge group. The action \eqref{ym1}, although not gauge invariant, is invariant under BRST transformations, namely,
\begin{eqnarray}
sY^\mathsf{a}&=&\nabla^\mathsf{ab}c_\mathsf{b}\;,\nonumber\\
sc^\mathsf{a}&=&\frac{\kappa}{2}f^\mathsf{abc}c_\mathsf{b}c_\mathsf{c}\;,\nonumber\\
s\overline{c}^\mathsf{a}&=&b^\mathsf{a}\;,\nonumber\\
sb^\mathsf{a}&=&0\;,\label{brst0}
\end{eqnarray}
where $s$ is the nilpotent BRST operator.

Let us make some important remarks about Yang-Mills theories.
\begin{itemize}

\item Although the fundamental fields and the action \eqref{ym1} are not gauge invariant, physical observables must be gauge invariant. This statement is perhaps the most sacred principle in gauge theories, it is called \emph{gauge principle}. This principle can be extended to BRST invariant operators in order to include more general operators that might be related to physical quantities. At classical level, the BRST invariance principle reduces to the usual gauge principle.

\item Together with quantum electrodynamics \cite{Tomonaga:1946zz,Schwinger:1948yk,Feynman:1950ir}, and the Higgs mechanism \cite{Higgs:1964pj,Higgs:1964ia}, Yang-Mills theories are the theoretical basis of the Standard model. Thus, it is a successful theory with a deep connection with reality.

\item The action \eqref{ym1} is a renormalizable theory, at least to all orders in perturbation theory \cite{'tHooft:1972fi,Piguet:1995er}. Renormalizability states that ultraviolet divergences can be consistently eliminated from perturbative computations, \emph{i.e.}, Yang-Mills theories are stable at quantum level. It is worth mention that BRST symmetry is very important with respect to the renormalizability of a gauge theory. 

\item Another feature of the action \eqref{ym1} is unitarity \cite{Kugo:1979gm}. It states that the $S$-matrix conserves probability and, thus, asymptotic states can be defined. For pure gauge theories, unitarity requires a well defined BRST symmetry and that the gauge group is compact.

\item The renormalization of the coupling parameter leads to the concept of asymptotic freedom \cite{Gross:1973id,Politzer:1973fx}. This property predicts that, at high energies, the coupling is very small and perturbation theory can be safely employed. However, as the energy decreases the coupling increases. At this strong coupling regime, the theory is non-perturbative and yet to be fully understood.

\item Confinement \cite{Baulieu:2009xr,Dudal:2012sb,Kugo:1979gm} states that quarks and gluons cannot be observable states. Actually, the physical observables must be, not only gauge invariant, but also colorless. At high energies, confinement manifests itself as the quark-gluon plasma, a state where quarks and gluons are almost free. However, they cannot be distinguished from the plasma. At low energies, the strong coupling enforces hadronization phenomena, \emph{i.e.}, the physical spectrum is composed by hadrons and glueballs. Although experimentally established, it is a theoretical challenge to show that the Yang-Mills action leads to an effective theory composed by hadrons and glueballs.

\item Another possible effect is that, at low energies, dynamical mass generation \cite{Dudal:2005na,Dudal:2011gd} takes place. It originates from the condensation of dimension-2 operators. The mass parameters that emerge drastically change the behavior of the theory at infrared scale and are relevant for confinement.

\item It was shown in \cite{Gribov:1977wm,Singer:1978dk} that, at low energies, gauge fixing is not possible by the simple introduction of a constraint. This problem is known as Gribov ambiguities problem. The main point in the elimination of the Gribov ambiguities is that BRST quantization (and also the standard Faddeev-Popov quantization \cite{Faddeev:1967fc}) shows itself to be incomplete at low energy level. In fact, the implementation of a gauge fixing does not eliminate completely the gauge symmetry \cite{Gribov:1977wm}, a residual symmetry survives. Moreover, it was shown that this problem  occur for any gauge choice \cite{Singer:1978dk}. The spurious gauge configurations are called Gribov copies and they are the kernel of the Gribov ambiguities. Remarkably, these copies gain relevance only at low energies, keeping the high energy sector untouched. To eliminate the Gribov ambiguities is a hard step that is not yet fully understood. However, at the Landau gauge, the inifinitesimal copies can be eliminated through the introduction of a soft BRST breaking\footnote{A soft breaking is breaking whose field operators have dimension lower than the spacetime dimension.} related to a mass parameter known as Gribov parameter. Outstandingly, the treatment of this technical problem leads to exceptional evidences of quark-gluon confinement \cite{Zwanziger:1992qr,Baulieu:2008fy,Baulieu:2009xr,Dudal:2012sb,Dudal:2005na}.

\end{itemize}

\subsection{The infrared puzzle, Gribov ambiguities and BRST soft breaking}\label{ir}

Basically, at the infrared regime, there are several techniques leading to confinement evidences. For instance, non-trivial solutions of the Schwinger-Dyson equations \cite{Binosi:2011zz}, lattice simulations \cite{Greensite:2003bk,Cucchieri:2011ig}, AdS/QCD duality \cite{Maldacena:1997zz,Witten:1998qj,Miranda:2009qp}, dual superconductivity and defects \cite{Nambu:1974zg,'tHooft:1975,Mandelstam:1974pi,Faddeev:2001dda} and dimension-2 condensates and Gribov ambiguities \cite{Gribov:1977wm,Zwanziger:1992qr,Sobreiro:2005ec,Baulieu:2008fy,Baulieu:2009xr,Dudal:2012sb,Dudal:2005na,Dudal:2011gd}. The puzzle consists in merging all evidences and formalisms in order to obtain a consistent theory in terms of hadrons and glueballs. As mentioned, we confine ourselves to explore the Gribov problem and its relation to confinement.

The treatment of Gribov ambiguities has the effect of changing the action \eqref{ym1} to the so called Gribov-Zwanziger action,
\begin{equation}
S=S_0+\int\left[\overline{\phi}_\mathsf{ac}\mathrm{d}\ast\nabla^\mathsf{ab}\phi_\mathsf{bc}-\overline{\omega}_\mathsf{ac}\mathrm{d}\ast\nabla^\mathsf{ab}\omega_\mathsf{bc}+\gamma^2f^\mathsf{abc}Y_\mathsf{a}\left(\overline{\phi}+\phi\right)_\mathsf{bc}+4M\gamma^4\right]\;,\label{ym2}
\end{equation}
where $M=\dim{G}$, $\gamma$ is the Gribov parameter which has dimension of a mass, the fields $\overline{\phi}_\mathsf{ab}$ and $\phi_\mathsf{ab}$ are 1-form bosonic fields while $\overline{\omega}_\mathsf{ab}$ and $\omega_\mathsf{ab}$ are fermionic 1-form fields. These fields obey the following transformation rules,
\begin{eqnarray}
s\overline{\omega}_\mathsf{ab}&=&\overline{\phi}_\mathsf{ab}\;,\nonumber\\
s\overline{\phi}_\mathsf{ab}&=&0\;,\nonumber\\
s\phi_\mathsf{ab}&=&\omega_\mathsf{ab}\;,\nonumber\\
s\omega_\mathsf{ab}&=&0\;,\label{brst1}
\end{eqnarray}
and can be eliminated through field equations in favor of a nonlocal term, the horizon-function \cite{Zwanziger:1992qr,Dudal:2005na,Dudal:2011gd}.

We now establish a few properties of the action \eqref{ym2}.
\begin{itemize}

\item The action \eqref{ym2} is renormalizable, at least to all orders in perturbation theory, in such a way that the ultraviolet sector of the theory is unchanged. In fact, no extra renormalizations are required \cite{Zwanziger:1992qr,Dudal:2005na}; all extra fields and Gribov parameter renormalization factors depend on the gluon and coupling parameter renormalization factors.

\item The action \eqref{ym2} is not BRST invariant, and the breaking is proportional to the Gribov parameter, $sS\propto\gamma^2$. Since $\gamma$ has dimension of a mass, the breaking is soft, and thus, harmless to the UV sector \cite{Baulieu:2008fy,Baulieu:2009xr}.

\item The Gribov parameter is determined from the minimization of the quantum action $\delta \Sigma/\delta \gamma^2=0$. A simple computation at the tree level predicts that $\gamma^2=\mu^2\exp\{-\frac{64\pi^2}{3N\kappa^2}\}$, where $\mu^2$ is a cutoff and $N$ is the Casimir of the gauge group, $f^\mathsf{acd}f_\mathsf{bcd}=-N\delta^\mathsf{a}_\mathsf{b}$. Thus, at the perturbative regime, $\kappa\rightarrow0$, we have that $\gamma\rightarrow0$. To take this limit is equivalent to take the high energy limit. Thus, the presence of the Gribov parameter allows the action to be continuously deformed into action \eqref{ym1}. Then, the UV and IR sectors can be continuously deformed to each other. Moreover, the BRST is asymptotically restored at the UV limit.

\item One of the evidences of confinement emerges from the gauge propagator which, at finite values of $\gamma$, acquires imaginary poles. The consequence is that it violates positivity of the spectral representation of this propagator and thus, no physical particles can be associated with this propagator \cite{Dudal:2005na}. 

\item More recently \cite{Dudal:2011gd}, the union of the Gribov-Zwanziger formalism and the condensation of dimension-2 operators has been developed. This is called the \emph{refined Gribov-Zwanziger formalism} and improves the results obtained from the Gribov-Zwanziger action. For instance, the propagators of the RGZ action coincide with the results obtained from lattice simulations \cite{Cucchieri:2011ig}.

\item We also remark that, some advances on the determination of the physical spectrum of the infrared sector of Yang-Mills theories have been made \cite{Dudal:2010cd,Capri:2012hh}. Although very difficult, the main requirement is gauge invariance, \emph{i.e.}, observables are related with gauge invariant operators. In QCD, these operators must describe its low energy spectrum, \emph{i.e.}, hadrons and glueballs.

\end{itemize}

\section{QCD-Gravity analogy}\label{qcdgrav}

This work is based on the idea that gravity can be described at quantum level by a gauge theory in a Euclidean four-dimensional spacetime. Classical gravity emerges if one could possibly show that a geometrodynamical theory rises at the infrared sector. This is performed by applying the soft BRST breaking technique. Let us elaborate this idea in more detail.

We assume that quantum gravity can be described by a Yang-Mills theory in four-dimensional Euclidean spacetime, $\mathbb{R}^4$. This choice, instead of a Minkowski spacetime, has three main reasons: (i) first, because QFT is only solvable in Euclidean spaces. Reliable computations require a Wick rotation from the Minkowkian to the Euclidean space \cite{Itzykson:1980rh}. However, a Wick rotation is known to be valid only at perturbative level. At non-perturbative scales, there are no indications that a Wick rotation can be performed. Thus, since we are interested in non-perturbative effects, it is necessary to start with a Euclidean spacetime. (ii) The second reason, and perhaps the most important, is that the Euclidean space is the simplest geometry. Since we are constructing a theory that will determine spacetime geometry, to start with the simplest geometry is the right choice. (iii) In four dimensions, the coupling parameter of a Yang-Mills theory is dimensionless. As a consequence, pure Yang-Mills action is massless\footnote{In Yang-Mills theories, the four-dimensional case is the only case where the coupling parameter is dimensionless.}. This fact is important because it prevents the gauge field, which has dimension 1, to be associated with the vierbein, which has vanishing dimension.

The gauge field is an algebra-valued 1-form and thus, carries $\dim{G}$ components with respect to the algebra of the gauge group. The gauge group $G$ has to fulfill a few requirements \cite{Sobreiro:2011hb,Sobreiro:2012book}: (i) the first one is that $\dim{G}\ge\dim{ISO(4)}$, in such a way that the degrees of freedom of gravity are covered. (ii) The universal principal bundle that describes this theory must be non-trivial, so the Gribov problem is inherent to the theory and has to be properly treated \cite{Singer:1978dk,Daniel:1979ez,Nakahara:1990th}. (iii) The group must decompose at least\footnote{Larger groups are also allowed. The extra decompositions will generate a matter sector in the resulting gravity theory, \cite{Assimos:2012zzz}.} as $G=H+Q$ where $Q=G/H$ must be a symmetric space and $H$, obviously, a stability group. The stable group must share a morphism with Lorentz-type groups and $Q$ must define a vector representation of $H$. Thus, $H$ can be identified with local Lorentz transformations and $Q$ with a sector that expand the vierbein. As a consequence, the gauge components can be identified with the spin-connection and the vierbein. (iv) It can be very convenient to have a symmetry breaking $G\rightarrow H$, in such a way that the field at the $Q$ sector acquires a matter-type transformation with respect to $H$. That is why $Q$ must be a symmetric space. It is important to mention that, depending on the gauge group, a symmetry breaking might not be needed. This occurs if the gauge transformations could be partially identified with local Lorentz transformations and partially with diffeomorphisms. A renowned example can be found in \cite{Witten:1988hc}.

It is important that the theory develops at least one mass parameter, possibly the Gribov parameter, in order to rescale the gauge field at the sector $Q$. As mentioned, this field, being a connection, carries dimension 1 and thus, cannot be directly associated with the vierbein. Only after the emergence of these mass parameters is that this identification can be performed. Moreover, the Gribov parameter and the consequent BRST breaking are important concepts that must be interpreted as follows: 

\emph{The theory has two sectors, the UV, which is a massless gauge theory in Euclidean four-dimensional space, and the IR sector, which presents soft BRST symmetry breaking and dynamically generated mass parameters. The UV sector is a standard, non-Abelian, asymptotic free, gauge theory of spin-1 excitations. Although the degrees of freedom coincide in number with a first order gravity, this identification is forbidden unless a mass parameter arises, so that a vierbein can be defined. Once the energy starts to decrease, the soft BRST symmetry breaking takes place, the Gribov parameter and other possible masses appear. At this stage, propagators of the fundamental fields develop complex poles, a fact that is interpreted as an evidence that these excitations are ruled out from the physical spectrum of the theory (in QCD, this is recognized as confinement). Physical observables must be defined at this point. In QCD, the observables are hadrons and glueballs. In gravity, the low energy observables must be \underline{geometry}. Thus, one possibility is to identify the $H$ gauge field with the spin-connection and the $Q$ field with the vierbein. The consequence is a first order gravity where geometry is determined by the usual relations \eqref{geom0} and \eqref{geom1}.}

If all above requirements are fulfilled, we gain a deep analogy between gravity and quantum chromodynamics. At high energies, both theories are well defined quantum gauge theories in a four-dimensional flat space. Both theories present soft BRST breaking and have to be redefined in order to establish the physical content at the infrared regime. In the case of QCD, the physical content are confined states identified with hadrons and glueballs. In gravity, the physical content are identified with geometric properties of spacetime.

We now turn to a specific model when these ideas are realized.

\section{Realization of the idea through de Sitter groups}\label{realiz}

\subsection{de Sitter-Yang-Mills theory}

One possible realization of the above ideas occurs in a gauge theory based on the group $SO(m,n)$ with $m+n=5$ and $m\in\{0,1,2\}$. For  $m=0$, the gauge group is the orthogonal one. For $m=1$ and $m=2$, we have a de Sitter and anti de Sitter groups, respectively\footnote{Except when necessary, we will indistinguishably call the generic $SO(m,n)$ group, with arbitrary $m$, by de Sitter group.}. Spacetime is a Euclidean four-dimensional differential manifold $\mathbb{R}^4$. The algebra of the group is given by
\begin{equation}
\left[J^{AB},J^{CD}\right]=-\frac{1}{2}\left[\left(\eta^{AC}J^{BD}+\eta^{BD}J^{AC}\right)-\left(\eta^{AD}J^{BC}+\eta^{BC}J^{AD}\right)\right]\;,\label{alg1}
\end{equation}
where $J^{AB}$ are the $10$ anti-hermitian generators of the gauge group, antisymmetric in their indices. Capital Latin indices are chosen to run as $\{5,0,1,2,3\}$. Comparing with the generic gauge theory described in Sect.~\ref{gauge}, the indices identification is $\mathsf{a}\equiv AB=-BA$. The $SO(m,n)$ group defines a five-dimensional flat space, $\mathbb{R}^{m,n}_S$, with invariant Killing metric given by $\eta^{AB}\equiv\mathrm{diag}(\epsilon,\varepsilon,1,1,1)$ with $\epsilon=(-1)^{(2-m)!}$ and $\varepsilon=(-1)^{m!+1}$. The spaces $\mathbb{R}^4$ and $\mathbb{R}^{m,n}_S$ have no dynamical relation whatsoever.

The $SO(m,n)$ Yang-Mills action is renormalizable and is defined over a non-trivial universal bundle \cite{Sobreiro:2011hb,Sobreiro:2012book,Assimos:2012zzz}. Thus, BRST soft breaking takes place. Moreover, a few extra mass parameters may emerge. The presence of these masses will be forwardly used. However, we will fix our attention only at the Yang-Mills action and the gauge field $Y$.

The de Sitter group may be decomposed as a direct product, $SO(m,n)\equiv SO(m!-1,n)\otimes S(4)$ where $S(4)\equiv SO(m,n)/SO(m!-1,n)$ is a symmetric coset space with four degrees of freedom. This decomposition is carried out by projecting the group space in the fifth coordinate $A=5$.
 Defining then $J^{5a}=J^a$, where small Latin indices vary as $\{0,1,2,3\}$, the algebra \eqref{alg1} decomposes as
\begin{eqnarray}
\left[J^{ab},J^{cd}\right]&=&-\frac{1}{2}\left[\left(\eta^{ac}J^{bd}+\eta^{bd}J^{ac}\right)-\left(\eta^{ad}J^{bc}+\eta^{bc}J^{ad}\right)\right]\;,\nonumber\\
\left[J^a,J^b\right]&=&-\frac{\epsilon}{2}J^{ab}\;,\nonumber\\
\left[J^{ab},J^c\right]&=&\frac{1}{2}\left(\eta^{ac}J^b-\eta^{bc}J^a\right)\;,\label{alg2}
\end{eqnarray}
where $\eta^{ab}\equiv\mathrm{diag}(\varepsilon,1,1,1)$.

The gauge connection, follows the same decomposition, $Y=Y^A_{\phantom{A}B}J_A^{\phantom{A}B}=A^a_{\phantom{a}b}J_a^{\phantom{a}b}+\theta^aJ_a$ and the gauge transformations in Eq.~\eqref{gt0} are decomposed, at infinitesimal level, as
\begin{eqnarray}
 {A}^a_{\phantom{a}b}&\longmapsto& {A}^a_{\phantom{a}b}+\mathrm{D}\alpha^a_{\phantom{a}b}-\frac{\epsilon\kappa}{4}\left(\theta^a\xi_b-\theta_b\xi^a\right)\;,\nonumber\\
\theta^a&\longmapsto&\theta^a+\mathrm{D}\xi^a+\kappa\alpha^a_{\phantom{a}b}\theta^b\;.\label{gt2}
\end{eqnarray}
where the full gauge parameter splits as $\zeta=\alpha^a_{\phantom{a}b}J_a^{\phantom{a}b}+\xi^aJ_a$ and $\mathrm{D}=\mathrm{d}+\kappa A$ is the covariant derivative with respect to the sector $SO(m!-1,n)$. The field strength also decomposes, $F=\left(\Omega^a_{\phantom{a}b}-\frac{\epsilon\kappa}{4}\theta^a\theta_b\right)J_a^{\phantom{a}b}+K^aJ_a$, where $\Omega^a_{\phantom{a}b}=\mathrm{d}A^a_{\phantom{a}b}+\kappa A^a_{\phantom{a}c}A^c_{\phantom{c}b}$ and $K^a=\mathrm{D}\theta^a=\mathrm{d}\theta^a-\kappa A^a_{\phantom{b}b}\theta^b$. Under this decomposition, the Yang-Mills action \eqref{ym0} is written as
\begin{equation}
S=\frac{1}{2}\int\left[\Omega^a_{\phantom{a}b}{*}\Omega_a^{\phantom{a}b}+\frac{1}{2}K^a{*}K_a-\frac{\epsilon}{2}\Omega^a_{\phantom{a}b}{*}(\theta_a\theta^b)+\frac{1}{16}\theta^a\theta_b{*}(\theta_a\theta^b)\right]\;.\label{ym0a}
\end{equation}

Once the energy starts to decrease the set of mass parameters, together with BRST soft breaking, dynamically arises. At this point, it is convenient to perform the following redefinitions\footnote{The transformations \eqref{resc1} are not accidental.
 Both sectors are rescaled with $\kappa^{-1}$ in order to factor out the coupling parameter outside the action, a standard procedure in Yang-Mills theories \cite{Itzykson:1980rh}.
 On the other hand, the mass parameter affects only the $\theta$-sector, transforming it in a field with dimensionless components.
 It turns out that this is the unique possibility if one wishes to identify $\theta$ with a vierbein field.
 If also $A$ is rescaled with a mass factor, then it would never be possible to identify it with the spin connection.}
\begin{eqnarray}
A&\longmapsto&\kappa^{-1}A\;,\nonumber\\
\theta&\longmapsto&\kappa^{-1}m\theta\;,\label{resc1}
\end{eqnarray} 
where $m$ is a mass scale depending on the mass parameters of the theory. The action \eqref{ym0a} is then rescaled to
\begin{equation}
S=\frac{1}{2\kappa^2}\int\left[\overline{\Omega}^a_{\phantom{a}b}{*}\overline{\Omega}_a^{\phantom{a}b}+\frac{m^2}{2}\overline{K}^a{*}\overline{K}_a-\frac{\epsilon m^2}{2}\overline{\Omega}^a_{\phantom{a}b}{*}(\theta_a\theta^b)+\frac{m^4}{16}\theta^a\theta_b{*}(\theta_a\theta^b)\right]\;,\label{ym1a}
\end{equation}
where $\overline{\Omega}^a_{\phantom{a}b}=\mathrm{d} {A}^a_{\phantom{a}b}+ {A}^a_{\phantom{a}c} {A}^c_{\phantom{c}b}$, $\overline{K}^a=\overline{\mathrm{D}}\theta^a$, and the covariant derivative is now $\overline{\mathrm{D}}=\mathrm{d}+A$. Moreover, a reparameterization of the $SO(m,n)$ generators is required due to the existence of a mass scale, \emph{i.e.}, a stereographic projection is now allowed if one identifies the mass parameter with the radius of the gauge manifold $\mathbb{R}^{m,n}_S$, see \ref{ap1}. The consequence for de Sitter algebra \eqref{alg1} is
\begin{eqnarray}
\left[J^{ab},J^{cd}\right]&=&-\frac{1}{2}\left[\left(\eta^{ac}J^{bd}+\eta^{bd}J^{ac}\right)-\left(\eta^{ac}J^{bc}+\eta^{bc}J^{ad}\right)\right]\;,\nonumber\\
\left[J^a,J^b\right]&=&-\frac{\epsilon m^2}{2\kappa^2}J^{ab}\;,\nonumber\\
\left[J^{ab},J^c\right]&=&\frac{1}{2}\left(\eta^{ac}J^b-\eta^{bc}J^a\right)\;.\label{alg3}
\end{eqnarray}

\subsection{Contraction and symmetry breaking}\label{cont}

The main trick, in order to obtain gravity, is to consider that the rate $m^2/\kappa^2$ tends to vanish at low energy scales\footnote{We remark that, due to the hierarchy of fundamental interactions, for gravity, a low energy scale would be very high when compared to all other basic interactions. The rate $m^2/\kappa^2$ may, become small again at lower scales.}. Due to asymptotic freedom, this is a very consistent hypothesis. Thus, we can perform an In\"on\"u-Wigner contraction \cite{Inonu:1953sp} through $m^2/\kappa^2\rightarrow 0$, at the algebra \eqref{alg3}, enforcing the de Sitter group to be contracted down to the Poincar\'e group. The second of relations \eqref{alg3} deforms to $\left[P^a,P^b\right]=0$, where $J^a\longmapsto -\kappa\gamma^{-1}P^a$ and $\theta\longmapsto-\theta^aP_a$ (See the limit of \eqref{eq:thetaJa} in \ref{ap1}). The gauge symmetry is then dynamically deformed to the Poincar\'e group, $SO(m,n)\longrightarrow ISO(m!-1,n)$, for some values $\kappa$ in the strong coupling regime. The details of this deformation can be understood as a stereographic projection, see \ref{ap1}.

Typically, an In\"on\"u-Wigner contraction is a deformation of the group to another group which is not a subgroup of the former. Thus, since the Poincar\'e group is not a symmetry of the action \eqref{ym1a}, a dynamical symmetry breaking takes place. In fact, the Lorentz-type group $SO(m!-1,n)$ is a common subgroup, $ISO(m!-1,n)\supset SO(m!-1,n)\subset SO(m,n)$, and also a stability subgroup for both groups. Thus, the theory suffers a symmetry breaking $SO(m,n)\longrightarrow SO(m!-1,n)$. Under the $SO(m!-1,n)$ gauge symmetry,
 the transformations \eqref{gt2} reduce to
\begin{eqnarray}
 A^a_{\phantom{a}b}&\longmapsto& {A}^a_{\phantom{a}b}+\overline{\mathrm{D}}\alpha^a_{\phantom{a}b}\;,\nonumber\\
\theta^a&\longmapsto&\theta^a-\alpha^a_{\phantom{a}b}\theta^b\;,\label{gt3}
\end{eqnarray}
where \eqref{resc1} was assumed. Thus, at action \eqref{ym1a}, the field $A$ is a gauge field with respect to the Lorentz group while $\theta$ has migrated to the matter sector (it suffers only group rotations under infinitesimal Lorentz transformations, \emph{i.e.}, it is a vector representation). Now, the theory is ready to be identified with gravity.

\subsection{Effective gravity}\label{grav}

The broken theory described in Sect.~\ref{cont} can generate an effective geometry if \cite{Sobreiro:2011hb,Sobreiro:2012book,Assimos:2013zzz}: (i) every configuration $(A,\theta)$ defines an effective geometry $(\omega,e)$; (ii) there exists a mapping from each point $x\in\mathbb{R}^4$ to a point $X\in\mathbb{M}^4$ of the deformed space. In order to preserve the algebraic structure already defined in $\mathbb{R}^4$, it is demanded that this mapping is an isomorphism; (iii) The local gauge group $SO(m!-1,n)$ defines, at each point of the mapping, the isometries of the tangent space $T_X(\mathbb{M})$. Thus, $\theta$ and $A$ can be identified with the vierbein $e$ and spin connection $\omega$, respectively, through
\begin{eqnarray}
\omega^{\mathfrak{ab}}_\mu(X)dX^\mu&=&\delta^{\mathfrak{a}}_a\delta^{\mathfrak{b}}_bA^{ab}_\mu(x)dx^\mu\;,\nonumber\\
e^{\mathfrak{a}}_\mu(X)dX^\mu&=&\delta_a^{\mathfrak{a}}\theta^a_\mu(x)dx^\mu\;.\label{id2}
\end{eqnarray}
In expressions \eqref{id2}, latin indices $\{\mathfrak{a},\mathfrak{b}\ldots\}$ belong to the tangent space $T_X(\mathbb{M})$. Moreover, it is always possible to impose that the space of all $p$-forms in $\mathbb{R}^4$ is mapped into the space of $p$-forms in $\mathbb{M}^4$, namely, $\Pi^p\longmapsto\widetilde{\Pi}^p$, and the same for the Hodge spaces, $\ast\Pi^p\longmapsto\star\widetilde{\Pi}^p$, where $\star$ is the Hodge dual in $\mathbb{M}^4$. This mapping, together with the identifications in Eq.~\eqref{id2}, provides
\begin{equation}
S=\frac{1}{8\pi G}\int\left[\frac{1}{2\Lambda^2}R^\mathfrak{a}_{\phantom{a}\mathfrak{b}}\star R_\mathfrak{a}^{\phantom{a}\mathfrak{b}}+T^\mathfrak{a}\star T_\mathfrak{a}-\frac{\epsilon}{2}\epsilon_\mathfrak{abcd}R^\mathfrak{ab}e^\mathfrak{c}e^\mathfrak{d}+\frac{\Lambda^2}{4}\epsilon_\mathfrak{abcd}e^\mathfrak{a}e^\mathfrak{b}e^\mathfrak{c}e^\mathfrak{d}\right]\;,\label{ym3}
\end{equation}
where $R^\mathfrak{a}_{\phantom{a}\mathfrak{b}}=\mathrm{d}\omega^\mathfrak{a}_{\phantom{a}\mathfrak{b}}+\omega^\mathfrak{a}_{\phantom{a}\mathfrak{c}} \omega^\mathfrak{c}_{\phantom{c}\mathfrak{b}}$ and $T^\mathfrak{a}=\mathrm{d}e^\mathfrak{a}-\omega^\mathfrak{a}_{\phantom{a}\mathfrak{b}}e^\mathfrak{b}$ are the curvature and torsion, respectively. Newton and cosmological constants are determined by $m^2=\kappa^2/2\pi G$ and $\Lambda^2=m^2/4$.

Action \eqref{ym3} is a gravity action in the first order formalism presenting (in order of appearance) a quadratic Yang-Mills-type curvature term, a quadratic torsion term, the Einstein-Hilbert term, and the cosmological constant term. Moreover, it is easy to check that the vacuum solution is a de Sitter-type spacetime with $T^\mathfrak{a}=0$ and $R^\mathfrak{ab}=2\epsilon\Lambda^2 e^\mathfrak{a}e^\mathfrak{b}$.

It is remarkable that, in the present theory, Newton and cosmological constants are related through $\Lambda^2=\kappa^2/8\pi G$. Obviously, from asymptotic freedom, $\kappa^2$ is a big quantity. And, by assumption, $G$ must be small. Thus, $\Lambda$ should be very big. In fact, if this is true, we can make two important remarks: (i) The first term in \eqref{ym3} can be safely neglected. Moreover, due to the absence of matter fields, torsion can be taken as very small. The resulting theory is then, the usual Einstein-Hilbert theory with cosmological constant; (ii) although astrophysical predictions \cite{Perivolaropoulos:2008pg,Padmanabhan:2012gv} determine that $\Lambda^2_{obs}$ is very small, quantum field theory predicts \cite{Nelson:1982kt,Toms:1983qr,Buchbinder:1986gj,Parker:1985kc} a very large $\Lambda^2_{qft}$. Thus, the contribution of a pure gravitational cosmological constant, which is big in our case, can drive the cosmological puzzle to a final consistent answer. In fact, following \cite{Sobreiro:2011hb,gr-qc/0611055,Shapiro:2009dh}, the renormalized cosmological constant of our model could be determined through $\Lambda^2_{ren}=\Lambda^2_{obs}-\Lambda^2_{qft}$.

\section{Some consistency checks}

\subsection{The explicit mapping}\label{map}
 
The formal aspects of the mapping between a gauge theory in Euclidean spacetime and a gravity theory can be discussed in terms of fiber bundle theory. The details can be found in \cite{Sobreiro:2011hb,Sobreiro:2012book,Assimos:2013zzz}. In Sect.~\ref{qcdgrav}, it was demanded that the mapping between the original space $\mathbb{R}^4$ and the effective space $\mathbb{M}^4$ is an isomorphism. In fact, it can be also shown that this mapping is unique and has an inverse, which ensures the absence of ambiguities and thus, its isomorphic character. The details of the proof can also be found in \ref{ap2}.

Let us summarize this result. We suppose that, in the $d$-dimensional original manifold, the metric tensor is $g_{\mu\nu}(x)$, where $x$ are the set of coordinates at a point in this space. The effective metric tensor is defined as $\widetilde{g}_{\mu\nu}(X)$, where $X$ is the set of coordinates at a point in the target space. Thus, the matrix that defines the map $x\longmapsto X$ for all points in the effective manifold is given by
\begin{equation}
L_{\phantom{\nu}\mu}^\nu=\left(\frac{\tilde{g}}{g}\right)^{1/2d}\tilde{g}^{\nu\alpha}g_{\alpha\mu}\;.\label{eq13}
\end{equation}
Its inverse is given by
\begin{equation}
\left(L^{-1}\right)_{\phantom{\nu}\mu}^\nu=\left(\frac{g}{\tilde{g}}\right)^{1/2d}g^{\nu\alpha}\tilde{g}_{\alpha\mu}\;.\label{eq15}
\end{equation}
In \eqref{eq13} and \eqref{eq15}, $g=\det{g_{\mu\nu}}$ and $\widetilde{g}=\det{\widetilde{g}_{\mu\nu}}$ are taken as non-vanishing quantities. Thus, ambiguities are absent in this mapping.

\subsection{Weinberg-Witten theorems and emergent gravity}

In \cite{Weinberg:1980kq}, Weinberg and Witten established two very powerful theorems. Essentially, they forbid: (i) massless charged states with helicity $j>1/2$ which have a conserved Lorentz-covariant current and (ii) massless states with helicity $j>1$ which have conserved Lorentz-covariant energy-momentum tensor. 

At first sight, the first theorem would forbid gauge theories to exist since, at high energies, vector bosons and gluons are charged massless states which carry helicity $j=1$. However, the conserved gauge current associated with these states are not Lorentz covariant \cite{Weinberg:1980kq}. Similarly, the second theorem forbids spin-2 states with conserved Lorentz-covariant energy-momentum tensor to be defined. In traditional GR the metric tensor $g_{\mu\nu}$ is not a Lorentz covariant tensor since it is constructed over a Riemannian manifold. The linearized version of GR, on the other hand, is also a gauge theory and the same principles of the first case apply \cite{Weinberg:1980kq}. Moreover, the energy-momentum tensor of the graviton field identically vanishes.

The Weinberg-Witten theorem (ii) applies, however, to emergent gravity theories for which massless spin-2 states are generated as effective or composite states that could be associated with graviton excitations. Nevertheless, one may argue that the Weinberg-Witten theorem (ii) applies to the present mechanism. However, this is not the case here. First, the theory has a few mass parameters and the theorem holds for massless states only. Second, and more important, there are no spin-2 states in this model. The fields are identified with geometry and not with spin-2 composite fields. Gravity emerges as geometrodynamics and not as a field theory for spin-2 particles in flat space. In fact, the mapping discussed in Sects.~\ref{grav} and \ref{map} has been employed at classical level. Nevertheless, renormalizability establishes that the quantum action has the same form of its classical version. The difference between the classical and quantum actions lies on the fields and parameters, which are their respective renormalized versions. Hence, although each configuration $(A,\theta)$ can define a geometry $(\omega,e)$, the mapping should not be applied to each configuration in the path integral, but at the quantum action itself. In this way, the resulting geometrodynamical theory is obtained from the full dynamical content of the original gauge theory. Moreover, the identification of the renormalized fields are made with respect to geometric quantities and not with spin-2 states. The conclusion is that there is no violation of the Weinberg-Witten theorem.

Nevertheless, after the emergence of gravity as geometrodynamics, linearization is allowed for weak gravitation and the spin-2 states might also be considered. In that case, these states are classical states associated to propagating fields. Quantization of these states can be done since it is not the fundamental theory, but an effective theory. Clearly, this does not violate the theorems because it fits in the same category of GR, \cite{Weinberg:1980kq}.

\subsection{Unitarity and equivalence principle}\label{unit}

One of the most important features of $SU(N)$ gauge theories is unitarity \cite{Itzykson:1980rh}, a property that, among other features, follows from the compact character of the gauge group. In the case of the present theory, unitarity is only ensured for $m=0$. In that case, the resulting gravity theory is an $SO(4)$ local isometric gravity. The local Euclidean character of spacetime provides a kind of incomplete prediction of the equivalence principle because it lacks the split between space and time. On the other hand, if we set\footnote{The case $m=1$ provides a non-unitary theory and also a local Euclidean spacetime with no difference between space and time.} $m=2$ from the beginning, the resulting theory has $SO(1,3)$ local isometries, \emph{i.e.}, localy, spacetime is a Minkowski space. Thus, there is a split between space and time originated from the breaking that lead to the mapping. This is nothing else than the rising of the equivalence principle.

Thus, if on the one hand, we start with a unitary gauge theory, the resulting gravity is not exactly the desired one because space and time are still indistinguishable. On the other hand, by giving up unitarity, the split between space and time correctly emerges. To solve the paradox is just a mathematical problem that can be fixed by imposing a Wick rotation during the mapping between $\mathbb{R}^4$ and $\mathbb{M}^4$. See \cite{Sobreiro:2011hb}.

Another argument can be provided by simply saying that unitarity is irrelevant because quantum gravity is far beyond Planck scale. Above Planck scale gravity is already a classical phenomenon. Moreover, due to confinement of Higgless non-Abelian gauge theories (such as QCD and the present gravity model), unitarity might be an overestimated property, mainly because the high energy state is a plasma with no observable singlet whatsoever (in the case of QCD, this state is the quark-gluon plasma).

\section{Final considerations}\label{end}

We have discussed that is possible to make a deep analogy between quantum chromodynamics and gravity. This analogy is realized if gravity could be described, at quantum level, by a gauge theory in a Euclidean four-dimensional spacetime while, at classical level, it is deformed to a geometrodynamical theory \cite{Sobreiro:2011hb,Sobreiro:2012book}. We have established the conditions for this mechanism to be realized and an example based on de Sitter-type groups was provided. In this mechanism, gravity arises as an emergent geometric theory. The requirements are asymptotic freedom, dynamical mass generation, and BRST soft breaking due to Gribov ambiguities.

The starting action \eqref{ym0}, for the $SO(m,n)$ gauge groups, lives in a four-dimensional Euclidean space, and thus, at high energies, it is a well defined quantum theory of spin-1 asymptotic free states. As the energy decreases, mass generation takes place, as well as the Gribov parameter and soft BRST breaking. The consequence is that the propagators acquire complex poles and are ruled out from the physical spectrum of the theory. At this point, an In\"on\"u-Wigner contraction is assumed and a symmetry breaking $SO(m,n)\rightarrow SO(m!-1,n)$ drives the theory to a gravity theory described by the action \eqref{ym3}. This action defines a first order gravity which has Einstein-Hilbert and cosmological constant terms. To obtain this action, a mapping has to be imposed. The consistency of this mapping was shown and an improvement has been made in order to harmonize the concepts of unitarity and the emergence of the equivalence principle.

A remarkable feature of this mechanism is that Newton and cosmological constant could be explicitly computed from the usual QFT techniques. Moreover, there is a constraint between these constants, namely, $\Lambda^2\propto\kappa^2/G$. Since $G$ is supposedly small and $\kappa$ is a big quantity at low energies, it is then expected that $\Lambda$ is very big. This property can be combined with the predictions of QFT for the cosmological constant in order to provide a value that agrees with the observed values. Also, from the fact that $\Lambda$ is big and the absence of matter in this work, it is possible to conclude that the action \eqref{ym3} can be safely approximated to the Einstein-Hilbert action with cosmological constant.

Let us make a comparison between the present gravity theory and the standard model. Strong and electroweak interactions are described by gauge theories. At high energies, these theories are very similar (except for the gauge groups). At low energies, however, these theories tend to behave in very different ways. Electroweak interactions suffer spontaneous symmetry breaking through the Higgs mechanism, giving rise to perturbative electrodynamics and massive gauge bosons. On the other hand, quark-gluon confinement shows up in chromodynamics, and hadronization phenomena take place. Specifically, confinement and the gauge principle state that physical observables must be gauge invariant and colorless. These states are recognized as hadrons and glueballs. Now, if the present theory can describe gravity, then: (i) at high energies, gravity is a gauge theory which is very similar to the other fundamental interactions; (ii) at low energies, instead of hadrons and glueballs, the physical observables are identified with geometry, and spacetime itself is affected by this  theory. Thus, geometry appears as the low energy manifestation of gravity, in the same way that hadronization and spontaneous symmetry breaking are the low energy manifestations of chromodynamics and electroweak interactions.

Finally, for the moment, we can only say that a standard four-dimensional renormalizable Yang-Mills theory can generate a gravity theory at low energy regime. Obviously, many computations and tests must be performed before we recognize this theory (or some variation) as \emph{the} quantum gravity theory or only an academic exercise.

\section*{Acknowledgements}

R.~F.~S.~is grateful to the organizing committee of \emph{NEB 15 - Recent Developments in Gravity} for the opportunity to deliver this talk. Conselho Nacional de Desenvolvimento Cient\'{i}fico e Tecnol\'{o}gico\footnote{RFS is a level PQ-2 researcher under the program \emph{Produtividade em Pesquisa}, 304924/2009-1.} (CNPq-Brazil) and the Coordena\c{c}\~ao de Aperfei\c{c}oamento de Pessoal de N\'{\i}vel Superior (CAPES) are acknowledged for financial support.

\appendix

\section{Stereographic projection}\label{ap1}

The group $SO(m,n)$ defines a flat space, $\mathbb{R}_S^{m,n}$, with metric given by $\eta^{AB}\equiv\mathrm{diag}(\epsilon,\varepsilon,1,1,1)$. In the presence of a mass scale $m$, we can define the radius of this space by
\begin{equation}
 \mathcal{R} = \frac{2\epsilon}{\left(\epsilon\varepsilon\right)^{\frac{1}{2}}} \left(\frac{\kappa}{m}\right)\;,\label{eq:radius}
\end{equation}
in such a way that
\begin{equation}
 \eta_{AB} \xi^{A}\xi^{B} = \varepsilon \mathcal{R}^{2} =\eta_{ab}\xi^{a}\xi^{b} + \epsilon(\xi^{5})^{2}\;,\label{aB:etaAB}
\end{equation}
where $\xi^A$ are Cartesian coordinates in $\mathbb{R}_S^{m,n}$.
Thus, the stereographic projection is defined by
\begin{eqnarray} \label{aB:proj}
 \xi^{a} &=& n \overline{x}^{a}, \nonumber\\
 \xi^5 &=& \left(\epsilon\varepsilon\right)^{\frac{1}{2}}\mathcal{R}\left(1-2n\right)\;,
\end{eqnarray}
where $\eta_{ab} = diag\ (\varepsilon,1,1,1)$. In order to have a projection conformally flat, \emph{i.e.}, $g_{ab} = n^2 \eta_{ab} $, it is demanded that
\begin{equation}
 n=\frac{1}{\left(1+\frac{\varepsilon\sigma^2}{4\mathcal{R}^2} \right)}\;,
\end{equation}
where $\sigma^{2} = \eta_{ab}\overline{x}^a\overline{x}^b$.

The effect of the projection on the group generators can also be studied. The group generators can be written as
\begin{equation} \label{aB:ger}
 J^A_{\phantom{A}B} = \frac{1}{2}\left(\eta^{AC}\xi_{C}\partial_{B} - \eta^{BC}\xi_{C}\partial_{A}\right)\;.
\end{equation}
Thus, it is a straightforward computation to show that
\begin{eqnarray}
 J^a_{\phantom{a}b} &=&  \frac{1}{2} \left( \overline{x}^a P^b - \overline{x}^b P^a \right)\;,\nonumber\\
  J^a &=&- \frac{\kappa}{m} P^{a} + \frac{\epsilon}{16}\frac{m}{\kappa} \left(2 \overline{x}_a \overline{x}^b P_b +\sigma^2 P^a \right)\;,\label{aB:Jabproj}
\end{eqnarray}
where \eqref{aB:proj} and \eqref{eq:radius} were used. Thus, from the second of relations \eqref{aB:Jabproj}, we have
\begin{equation}
 \theta\longmapsto\kappa^{-1}m\theta=-\theta^aP_a+\frac{\epsilon}{16}\frac{m^2}{\kappa^2}\theta^a\left(2\overline{x}_a\overline{x}_bP^b + \sigma^2 P_a\right). \label{eq:thetaJa}
\end{equation}

\section{Explicit derivation of the mapping}\label{ap2}

Let us only show how the explicit mapping between the spaces $\mathbb{R}^4$ and $\mathbb{M}^4$ can be obtained. We have considered that
\begin{eqnarray}
\Pi^p&\longmapsto&\tilde{\Pi}^p\;,\nonumber\\
*\Pi^p&\longmapsto&\star\tilde{\Pi}^p\;.\label{mapfull}
\end{eqnarray}
For generality purposes, we assume a generic original metric $g_{\mu\nu}$ which, eventually, we can set as a Euclidean metric. The effective metric is denoted by $\tilde{g}_{\mu\nu}$. Moreover, we can also consider manifolds with an arbitrary dimension $d$. Obviously, a necessary extra condition is that both $g=|\det{g_{\mu\nu}}|$ and $\tilde{g}=|\det{\tilde{g}_{\mu\nu}}|$ are non-vanishing quantities. To find the explicit mapping, we apply the first of \eqref{mapfull} to a generic $p$-form, 
\begin{equation}
f_{\mu_1\ldots\mu_p}(x)dx^{\mu_1}\ldots dx^{\mu_p}=\tilde{f}_{\mu_1\ldots\mu_p}(X)dX^{\mu_1}\ldots dX^{\mu_p}\;,\label{eq1}
\end{equation}
where $x\in\mathbb{R}^d$ and $X\in\mathbb{M}^d$. From \eqref{eq1}, one easily obtain
\begin{equation}
f_{\mu_1\ldots\mu_p}(x)=L_{\phantom{\nu_1}\mu_1}^{\nu_1}\ldots L_{\phantom{\nu_p}\mu_p}^{\nu_p}\tilde{f}_{\nu_1\ldots\nu_p}(X)\;,\label{eq3}
\end{equation}
where $L_{\phantom{\nu}\mu}^\nu=\frac{\partial X^\nu}{\partial x^\mu}$. For the corresponding Hodge dual we have,
\begin{eqnarray}
& &\sqrt{g}\epsilon_{\mu_1\ldots\mu_p\nu_{p+1}\ldots\nu_d}f^{\mu_1\ldots\mu_p}(x)dx^{\nu_{p+1}}\ldots dx^{\nu_d}=\nonumber\\
&=&\sqrt{\tilde{g}}\epsilon_{\mu_1\ldots\mu_p\nu_{p+1}\ldots\nu_d}\tilde{f}^{\mu_1\ldots\mu_p}(X)dX^{\nu_{p+1}}\ldots dX^{\nu_d}\;,\label{eq4}
\end{eqnarray}
from which one can find that
\begin{equation}
f^{\mu_1\ldots\mu_p}=\left({\frac{\tilde{g}}{g}}\right)^{1/2}\left(\frac{L}{d}\right)^{d-p}\tilde{f}^{\mu_1\ldots\mu_p}\;,\label{eq8}
\end{equation}
with $L=L^\mu_{\phantom{\mu}\mu}$. A comparison of \eqref{eq3} and \eqref{eq8} leads to
\begin{equation}
f_{\mu_1\ldots\mu_p}=\left({\frac{\tilde{g}}{g}}\right)^{1/2}\left(\frac{L}{d}\right)^{d-p}\tilde{g}^{\nu_1\alpha_1}g_{\alpha_1\mu_1}\ldots\tilde{g}^{\nu_p\alpha_p}g_{\alpha_p\mu_p}\tilde{f}_{\nu_1\ldots\nu_p}\;.\label{eq10}
\end{equation}
Combining \eqref{eq10} and \eqref{eq3}, we achieve
\begin{equation}
L_{\phantom{\nu}\mu}^\nu=\left({\frac{\tilde{g}}{g}}\right)^{1/2p}\left(\frac{L}{d}\right)^{(d-p)/p}\tilde{g}^{\nu\alpha}g_{\alpha\mu}\;,\label{eq11}
\end{equation}
which is not valid for $p=0$. In that case it is easy to find that \eqref{mapfull} is valid only if
\begin{equation}
\left({\frac{\tilde{g}}{g}}\right)^{1/2}\left(\frac{L}{d}\right)^d=1\;.\label{0f}
\end{equation}
The constraint \eqref{0f} implies that
\begin{equation}
L_{\phantom{\nu}\mu}^\nu=\frac{d}{L}\;\tilde{g}^{\nu\alpha}g_{\alpha\mu}\;.\label{eq11a}
\end{equation}
The trace $L$ can now be calculated from \eqref{0f} or \eqref{eq11a} providing
\begin{eqnarray}
L&=&d\left({\frac{g}{\tilde{g}}}\right)^{1/2d}\;,\nonumber\\
L&=&d^{1/2}(\tilde{g}^{\mu\nu}g_{\mu\nu})^{1/2}\;,\label{trace}
\end{eqnarray}
respectively. The relations \eqref{trace} enforce the extra constraint
\begin{equation}
(\tilde{g}^{\mu\nu}g_{\mu\nu})^{1/2}=d^{1/2}\left({\frac{g}{\tilde{g}}}\right)^{1/2d}\;.\label{constx}
\end{equation}
As a consequence, we obtain the final expression for the transformation matrix, which is given by Eq.~\eqref{eq13}. We recall that the effective metric is computed from the gravity field equations, while the original metric is a given quantity\footnote{For the present case we actually have ($d=4$ and $g_{\mu\nu}=\delta_{\mu\nu}$) $
L_{\phantom{\nu}\mu}^\nu=\left(\tilde{g}\right)^{1/8}\tilde{g}^{\nu\alpha}\delta_{\alpha\mu}$.}. It turns out that the mapping \eqref{eq13} has an inverse, given by Eq.~\eqref{eq15}. The existence of the inverse ensures the non-degeneracy of the mapping.

\end{document}